\documentclass[lettersize,journal]{IEEEtran}
\usepackage{amsmath,amsfonts}
\usepackage{algorithmic}
\usepackage{algorithm}
\usepackage{array}
\usepackage[caption=false,font=normalsize,labelfont=sf,textfont=sf]{subfig}
\usepackage{textcomp}
\usepackage{stfloats}
\usepackage{url}
\usepackage{verbatim}
\usepackage{graphicx}
\usepackage{cite}
\usepackage{xcolor}
\usepackage{soul}
\usepackage{tablefootnote}
\begin{document}

\title{A High-Power Microwave Limiter Using A Self-Actuated Plasma-Based EIT Topology}

\author{Muhammad Rizwan Akram,~\IEEEmembership{Member, ~IEEE,} Abbas Semnani,~\IEEEmembership{Senior Member, ~IEEE,}
        % <-this % stops a space
\thanks{The authors are with the Department of Electrical Engineering and Computer Science, The University of Toledo, Toledo, Ohio 43606, USA. (email: muhammadrizwan.akram@utoledo.edu; abbas.semnani@utoledo.edu). This work was supported by the Office of Naval Research (ONR) under Grant N00014-21-1-2441.}% <-this % stops a space
}

\markboth{IEEE TRANSACTIONS ON Microwave Theory and Techniques,~Vol.~x, No.~xx, 2024}%
{Shell \MakeLowercase{\textit{et al.}}: A Sample Article Using IEEEtran.cls for IEEE Journals}


\maketitle

\begin{abstract}
This paper presents a novel metamaterial topology incorporating gas discharge tubes for high-power microwave protection. The design features two split ring resonators positioned side by side with their splits oriented orthogonally. When exposed to low-power microwaves, each split ring resonator induces a resonance that interacts to create a passband within a broad stopband, facilitated by a phenomenon known as electromagnetically induced transparency (EIT). At high power levels, the integrated gas discharge tubes become ionized, forming plasma that acts as a switch to eliminate the EIT window, thereby reinstating the stopband for protection. Several prototypes have been developed for S-band operation based on this concept. Analytical, numerical, and experimental results are in complete agreement. The proposed device demonstrates superior protection with lower insertion loss in the OFF mode and higher isolation in the ON mode. Its strong ability to handle high-power microwaves is achieved using plasma-based switches instead of diodes, providing a reasonable response time and a straightforward design that enables rapid prototyping. Additionally, the device demonstrates frequency and power threshold tunability, highlighting its versatility as a microwave protection device.
\end{abstract}


\section{Introduction}
\IEEEPARstart{H}{igh} intensity electromagnetic waves \cite{a0sch2004} can disrupt the communication systems, interfere with critical electronic equipment, and cause malfunctioning, especially when the conductor's length resonates with the incoming waves, causing power surges. Typically, the protection is achieved by shielding through metallic enclosures or using absorbers. However, both approaches affect the system's ability to communicate with the outside world. To circumvent these issues, receivers are usually placed outside the shielding mechanisms. However, high-power microwaves still cause issues, especially when the received signal is amplified, which can severely damage the communication equipment. 

Power limiters are the standard backend components employed after the receiver antenna for in-band protection against high-power microwaves. The desirable features of the power limiters are the introduction of low insertion and high isolation for low- and high-power signals, respectively. The common topologies for these backend power limiters are based on ferrite materials\cite{a1adam2013}, semiconductor components (i.e., diodes and transistors) \cite{a2yang2009}, MEMS \cite{a3shojaei2016}, and plasma\cite{a4ramesh2023,a5yang2008, a6semnani2016}. However, there are several other scenarios where the high-power microwave can couple with the circuitry and cause a malfunction in the critical equipment.

In that broad sense, metamaterials offer a front-end platform to effectively limit high-power microwaves. Numerous solutions based on active metasurfaces integrated with lumped elements, i.e., p-i-n diodes \cite{a16li2017}, varactors \cite{a17li2018}, resistors \cite{a18li2015}, field effect transistors \cite{a19li2017}, and MEMS \cite{a20coutts2008} etc. have been proposed for such protection. These active metasurfaces require sensors to detect the power threshold and utilize the sensor output to shield the target by switching the active elements through a DC biasing scheme. However, the HPM duration can be very short, and any switching delay may fail to provide timely protection. To enable rapid protection, passive and adaptive protection techniques are desired. 

Passive/adaptive metasurfaces, on the other hand, utilize the incoming energy to trigger the protection mode by inducing a power-dependent nonlinear response. These passive solutions can be categorized as impedance surface- and frequency selective surface (FSS)-based solutions. For example, the impedance surfaces in \cite{a9huang2022, a10zhang2021} employed four symmetrical p-i-n diodes placed in a waveguide to limit the high-power waves, in which the isolation increased with the number of unit cells. These impedance surfaces lower the cutoff frequency of the propagating mode by emulating the short patches to the larger ones through the conducting diodes. Similarly, a varactor diode-based impedance surface was proposed in \cite{a22luo2015}, which induced the gradual isolation with the power. In \cite{a21homma2022}, anisotropic impedance surfaces were utilized, and protection was achieved by redirecting the high-intensity energy through the anisotropic bandgap induced with diodes. Recently, a split ring resonator integrated with p-i-n diodes has been proposed in which bandgap is induced under high power when the diode is in the ON state \cite{a23yang2022}. So far, limited isolation has been realized with these impedance surface-based schemes.

For the FSS-based solutions, the typical approach employs non-resonant meta-atoms using conductive strips integrated with p-i-n diodes on a multi-layer board \cite{a72022self,a8yang2013}. These strips are shorted under high power, forming a resonator to stop a particular frequency band. The non-resonant cases usually require a high level of microwave power to short the diodes, as the electric field is typically low. The resonant meta-atom requires at least two metallic layers separated by the spacers to induce electric and magnetic resonances for ideal transmission, increasing the overall thickness \cite{a11cho2016,a12zhou2021}. A trade-off between thickness and the overall bandwidth is necessary for such metamaterials. The other way is to use complementary structures based on dipole resonances \cite{a13zhao2019}. However, they typically require at least two switching elements for each polarization, i.e., four switches per unit cell. Although the protection is achieved in a particular band, the passband is shifted to a higher frequency depending on the conductivity of the switching elements, as slots are not entirely blocked. Moreover, all these schemes are based on p-i-n diodes as suboptimal switches because of their initial leakage current and because they can only withstand a small amount of power. Therefore, they are not suitable for very high-power scenarios.

On the other hand, recently, plasma-based protection schemes have received more attention as they are naturally tailored toward high-power applications. In these schemes, hermetically sealed gas elements are typically integrated with resonant structures. The gas ionizes in response to a high electric field, acting as a high-power switch for protection. Several prototypes have been demonstrated, especially for backend protection, such as microstrip-based limiter \cite{a4ramesh2023,a26cross2013,a27pascaud2015} and cavity limiter \cite{a25semnani2018,a6semnani2016}. In the context of front-end application, a waveguide plasma limiter has also been proposed, which requires 40 kW of pulsed power to form plasma and merely provides 3 dB of isolation \cite{a24woo2023}.    

In this article, we propose a novel resonant meta-atom topology based on the unique physics of electromagnetically induced transparency (EIT) \cite{a14tassin2009, a15papasimakis2008} to open up a passband inside a wide stopband. Since the experimental demonstration of the EIT phenomenon in electromagnetics in 2008 \cite{ a15papasimakis2008} and 2009 \cite{a14tassin2009}, several EIT-based designs have been investigated, particularly for slow-wave applications in optics and photonics. Most EIT applications in the microwave regime have been limited to sensing \cite{a29meng2012, yan2019terahertz,chen2020noncontact}. Here, for the first time, we utilize meta-atoms based on an EIT topology for high-power microwave (HPM) protection by incorporating a gas discharge tube. The meta-atom structure achieves HPM protection by incorporating a gas discharge tube (GDT). The enclosed gas undergoes breakdown under high-power microwave illumination, forming plasma as a switch, disrupting the passband introduced through the EIT mechanism. As a result, high isolation is achieved throughout the passband, which is non-linearly enhanced owing to the incident power intensity. The plasma-based switching topology combined with the EIT mechanism outperforms existing designs due to its ultra-thin profile, high power handling capability, low insertion loss, high isolation, fast switching, and simple structure, requiring only a few switches per unit, as shown in Table \ref{tab1:comp}.

\section{Theory and Design}
Electromagnetically induced transparency is a quantum-level phenomenon observed at three-level atomic media in which the perturbation of the secondary beam with well-determined tuning modifies the optical response of the primary laser beam. An EIT is typically achieved through the superposition of the dark mode with the bright mode or through the destructive quantum interference between different excitation pathways of the excited states. This enables a transparent window in an otherwise opaque media \cite{a28limonov2017}. However, the requirements of the required conditions, such as extremely low temperatures and sophisticated equipment, make it challenging to apply in practical applications in the quantum domain. Recently, an electromagnetic analog of EIT has been proposed using a metamaterial platform where coupled resonators could achieve a similar performance under normal conditions \cite{a14tassin2009,a15papasimakis2008}. The sharp scattering response of EITs enables slow wave propagation and enhances the wave-matter interaction \cite{a29meng2012}. This phenomenon is highly desirable in the proposed high-power microwave protection, where high-power EM waves interact with a gas discharge tube, causing gas breakdown and plasma formation. The enhanced interaction between EM waves and plasma as a lossy material results in high isolation over a wide bandwidth.

\begin{figure}[!]
\centering
\includegraphics[width=0.95\linewidth]{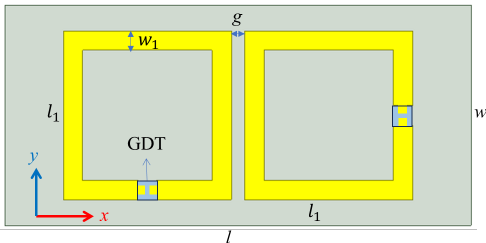}
\caption{Topology of the proposed microwave limiter based on an EIT configuration. Integrated gas discharge tubes (GDTs) between the split of the two split ring resonators exhibit power-dependent characteristics.}
\label{Fig1: topology}
\end{figure}

To experimentally implement an EIT-based microwave limiter, two split ring resonators are employed and placed side by side, as shown in Fig.~\ref{Fig1: topology}, so that the splits are orthogonal. Assuming a \textit{y}-polarized incidence, the SRRs interact differently to give rise to the coupled resonances—the right SRR couples with the incoming microwaves through the aligned split, known as bright mode. The left SRR has a split along the x-direction and forms another resonance known as dark mode. This dark mode couples the bright mode, making a transparent window, which is the EIT phenomenon.

\subsection{An Equivalent Circuit Modeling}

The EIT response can be modeled with coupled resonators using either spring-bound masses oscillator or their electric counterparts, i.e., RLC oscillators \cite{a31li2015}. Both of these approaches require the presence of external sources of disturbance or power, which complicates the modeling process. Recently, a four-pole model based on the bridge configuration of the RLC resonator model was proposed to capture the effect of the two excitation pathways as desired for EIT response \cite{a32}. This two-port model demonstrates the EIT response for the proposed topology.

\begin{figure}[!]
\centering
\includegraphics[width=0.9\linewidth]{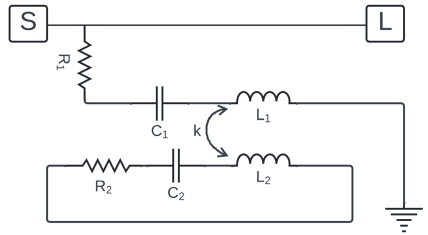}
\caption{An equivalent circuit model for the coupled resonators, capable of explaining the EIT response for a periodic surface.}
\label{Fig3: cir_mod}
\end{figure}

When the proposed unit cell is excited with a y-polarized plane wave, the left SRR acts as a ring resonator while the right SRR couples the incident wave through its split owing to its alignment with the incident wave, inducing two nearby resonant frequencies. This allows the extraction of RLC values from a single SRR with x- and y-polarized incidence using the FSS circuit modeling approach\cite{a33}:
\begin{equation}\label{eq3:sig}
C = \frac{\omega_1 - {\omega_2}^2/\omega_1}{\omega_2 Z_{\omega_1}-\omega_1 \omega_2 Z_{\omega_2}},
\end{equation}

\begin{equation}\label{eq4:cond}
L = \frac{Z_{\omega_1} + 1/{\omega_1}C}{\omega_1}.
\end{equation}
Here, $\omega_1$ and $\omega_2$ are the two frequency points against which impedances $Z_{\omega_1}$ and $Z_{\omega_2}$ are obtained. When brought closer to each other, these resonant frequencies partially overlap and can induce mutual coupling. From this knowledge, the circuit model of Fig.~\ref{Fig3: cir_mod} provides an approximate interplay of the coupled resonances that form the EIT under low power. In the limiting operation, i.e., high power incident, both the resonators act at nearly the same resonant frequency and form a stopband, and could be described with $k$ = 0. The underlying physics is further studied with the help of field analysis.

\subsection{Full-Wave Field Analysis}
\begin{figure}[!]
\centering
\includegraphics[width=0.95\linewidth]{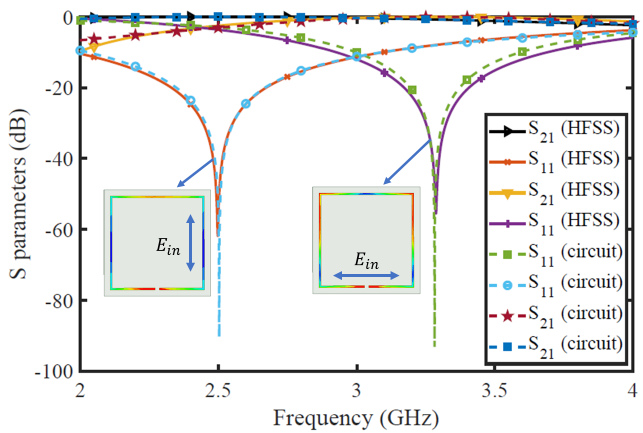}
\caption{Transmission responses of an SRR-based periodic surface under two different polarizations with L = 9.4 nH, and C = 0.42 pF for vertical incidence, and L = 16.8 nH, C = 0.14 pF for horizontal polarization.}
\label{Fig4: S22}
\end{figure}

\begin{figure}[!]
\centering
\includegraphics[width=0.95\linewidth]{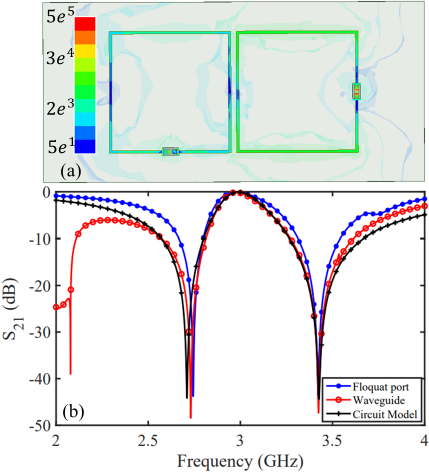}
\caption{The proposed topology under a low-power y-polarized TE incidence. (a) HFSS simulated $|E|$-field showing a strong field on the GDT placed vertically, enabling the high-power limiting mode. (b) Simulated transmission coefficients using analytical circuit model of Fig.~\ref{Fig3: cir_mod} ($R_1$ = 1 $\Omega$, $C_1$ = 0.165 pF, $L_1$ = 16.8 nH, $R_2$ = 0.1 $\Omega$, $C_2$ = 0.26 pF, $L_2$ = 10.9 nH, and k = 0.23), assuming source and load impedance of 377 $\Omega$, compared with HFSS simulation results using Floquat mode analysis and when placed in a WR-284 waveguide structure.}
\label{Fig2: S22}
\end{figure}

Suppose an SRR is illuminated by horizontal and vertical polarizations, as in Fig.~\ref{Fig4: S22}. In that case, each polarization excites a different mode, as confirmed by the E-field, resulting in a stopband at two neighboring frequencies. When these two resonances are excited simultaneously in the proposed unit cell topology, as in Fig.~\ref{Fig2: S22}(a), a transparent window opens accounting for the effects of EIT, as shown in Fig.~\ref{Fig2: S22}(b). The electric field in the split gap region of the right SSR over the electrodes of the GDT reaches a maximum of $5\times10^5$ V/m at 1 W of input power, as shown in Fig.~\ref{Fig2: S22}(a). As incident power increases, the gas eventually breaks down under such a strong E-field, and a plasma forms, which turns the SSR into a ring resonator and eliminates the transparent window.

To evaluate the EIT characteristics of the proposed topology, the Floquat mode periodic simulations are carried out using ANSYS High-Frequency Structure Simulator (HFSS) with primary/secondary boundary conditions. The transmission response for the y-polarized TE incident is presented in Fig.~\ref{Fig2: S22}. It is seen that a transparent window appears at a frequency of about 3 GHz, owing to the coupling between the two resonators. The proposed topology was designed considering the ease of characterization in the experimental setting for a WR-284 waveguide setup for high-power limiting verification. The transmission response is also evaluated when placed in the waveguide, and a good agreement is seen in Fig.~\ref{Fig2: S22}. Henceforth, only waveguide topology is pursued to simplify the high-power measurements.

\subsection{Power Limiting Performance}
Under high input power, the transparent window is converted to a stop band, causing plasma to be formed in the split gap region. The generated plasma acts as a switch and converts SSR to a complete ring. This ring resonator does not couple to the resonator on the left, eliminating the transparent window. Moreover, the plasma absorbs some power to sustain and dissipates some energy due to its Ohmic loss, further enhancing the isolation throughout the waveguide band for high-power microwave protection. 

Plasma is introduced in the gap region of the split ring resonator by using a commercially available GDT. The specific GDT utilized in this work is a hermetically sealed structure containing an inert gas at a pressure of 10-30 Torr between the two electrodes with a distance of about 400 $\mu$m. When the GDT is integrated with the SRR, its gap acts as the split of the SRR under low incident power. However, under high power, when the induced E-field across the electrodes surpasses the breakdown threshold, the gas inside the GDT ionizes, transitioning into a semiconductive plasma medium characterized by a complex permittivity, altering the SRR configuration to a ring resonator \cite{a30ramesh2021}:
\begin{equation}\label{eq1:sig}
\epsilon_{rp} = 1 - \frac{e^2n_e}{\epsilon_0 m\left(\omega^2+\nu_m^2\right)},
\end{equation}
\begin{equation}\label{eq2:cond}
\sigma_p = \frac{e^2n_e \nu_m}{m\left(\omega^2+\nu_m^2\right)}.
\end{equation}
Here, $e$, $m$, and $n_e$ are the electron charge, mass, and density, and $\epsilon_0$ is the free space permittivity. In addition, $\omega$ is the incident wave frequency, and $\nu_m$ is the electron-neutral collision frequency, which mainly depends on the composition and pressure of the enclosed gas. The plasma conductivity depends on the electron density, as seen in \ref{eq2:cond}, which is the function of the power delivered to the plasma. This conductivity eventually turns the resonator configuration from an SRR to a ring resonator, blocking the transparent window in the high-power mode.

\section{EIT-Based Limiter Devices}
To validate the proposed theory and topology, several prototypes are developed using Rogers TMM3 laminate ($\epsilon_r$ = 3.27 and tan$\delta$ = 0.002) with a thickness of 1.27 mm. The experimental characterizations are performed in a WR-284 waveguide system, which can be extended to free-space protection using an array of such unit cells.

\subsection{A Static Limiter}
\begin{figure}[!]
\centering
\includegraphics[width=0.95\linewidth]{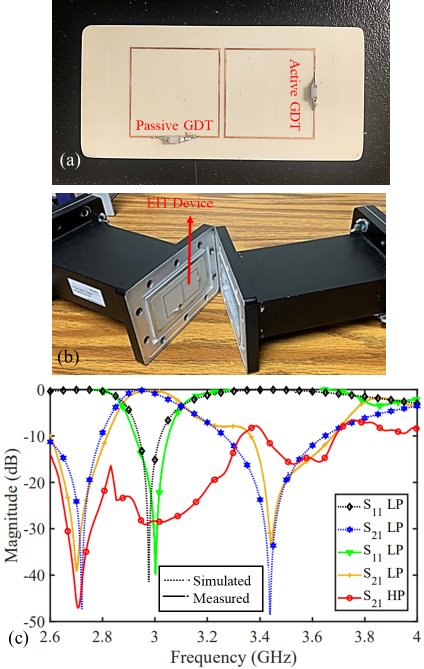}
\caption{(a) A fabricated prototype of the static EIT-based limiter with an active that turns on in response to high power and a passive GDT for structural symmetry, with \textit{l} = 72.14 mm, \textit{w} = 34.04 mm, $w_1$ = 0.5 mm, \textit{g} = 1 mm, and $l_1$ = 22 mm on a Rogers TMM3 laminate with a dielectric constant of 3.27 and $\tan\delta$ of $2\times10^{-3}$. (b) The WR-284 waveguide measurement setup and (c) Transmission/reflection coefficients of the static limiter under low (1 W) and high (100 W) input powers.}
\label{Fig5: Fab}
\end{figure}

The fabricated prototype of the static limiter is shown in Fig.~\ref{Fig5: Fab} (a). The dimensions are optimized for operation at 3 GHz and presented in the caption of Fig.\ref{Fig5: Fab}. Two commercially available GDTs, Littelfuse SE140 \cite{a34datasheet}, are used as plasma cells. GDTs are used on both SRRs to maintain structural symmetry. In the assembly process, the GDTs' electrodes are soldered to the ends of SRRs and form the resonators' gap.

The limiter device is placed inside a calibrated waveguide WR-284 measurement setup. The unit cell's outer dimensions are adjusted to fit inside one end of the waveguide and be sandwiched from the other side. The operating mode of the waveguide is TE$_{10}$ with a frequency band of 2.6 - 3.95 GHz. To evaluate the low-power performance and the EIT effect, an input power of 20 dBm (100 mW) is pushed into the waveguide, and the transmission coefficient, S$_{21}$, is measured. Numerically, the low power results are assumed by modeling plasma region using the Drude model for permittivity and conductivity, (\ref{eq1:sig}) and (\ref{eq2:cond}), with an assumed value of $n_e$ = 0. An explicit agreement between the numerical and experimental results is observed, as depicted in Fig.~\ref{Fig5: Fab}(b). In the plasma OFF (low power) mode, a 3-dB transmission bandwidth of 7.4\% from 2.86 to 3.08 GHz centered at 2.94 GHz with an insertion loss of 0.13 dB are observed in the simulations. In measurement, the passband window of 7.5\% ranging from 2.87 to 3.1GHz centered at 2.98 GHz with an insertion loss of 0.26 dB are achieved. A slight deviation from the simulation results is attributed to fabrication tolerances, manual assembly, and device placement inside the waveguide.  

To evaluate the high-power limiting performance, 50 dBm (100 W) of power is pushed into the waveguide structure. The measured high-power S$_{21}$ is plotted in red in Fig.~\ref{Fig5: Fab}(b). An isolation of more than 25 dB is achieved throughout the passband, and 8 dB of isolation is observed all over the waveguide band. Such high isolation in the passband is mainly attributed to the high plasma conductivity formed in the GDTs due to high input power, which eliminates the EIT response by short-circuiting the SRR's gap. The wideband isolation, on the other hand, is mainly attributed to the plasma absorptive and lossy properties.

\begin{figure}[!]
\centering
\includegraphics[width=0.95\linewidth]{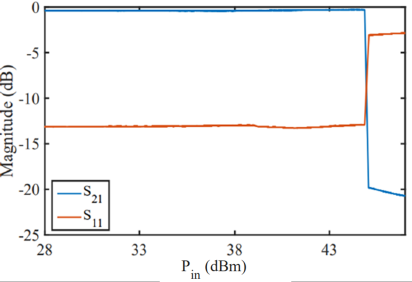}
\caption{Measured transmission and reflection responses of the static limiter against the input power at the passband center frequency of 3 GHz.}
\label{Fig5b: S11}
\end{figure}

To precisely evaluate the absorptive behavior of the introduced EIT limiter, the device is tested at the center frequency of 3 GHz with an input power sweep from 28 to 48 dBm. The measured transmission and reflection coefficients are shown in Fig.~\ref{Fig5b: S11}. Gas breakdown in the GDT happens at around 45 dBm of input power, and 20 dB of attenuation (isolation) is achieved owing to the plasma formation, which changes the SRR's characteristics. Immediate protection against high input powers is highly desirable for many applications, putting plasma-based protection at the forefront of the technologies compared to other solutions (e.g., semiconductors and MEMS). It should be noted that in the proposed topology, the limitation of the transmitted power is realized through a combination of power absorption and reflection. As seen in Fig.~\ref{Fig5b: S11}, even after gas breakdown and plasma formation, the measured reflection coefficient (S$_{11}$) remains at around 3 dB, showing that just about half of the input power is reflected. Considering more than 20 dB of isolation, the rest of the energy is absorbed and dissipated by the plasma, which gives a semi-absorptive character to such self-sustained plasma limiters.

The proposed plasma-based limiter has a long lifetime as it was designed for the GDT to operate in the diffusion-controlled regime above the critical frequency point \cite{Semnani:freq2013}. At this regime, heavy ions cannot reach the electrodes due to their mass and inability to follow the rapid transitions of the microwave field direction. This results in oscillatory ion behavior at the center of the gap, effectively preventing ion bombardment of the electrodes—a primary failure mechanism in plasma cells. In this regime, the influence of boundaries (electrodes) is minimal, and the discharge is primarily driven by electron-impact ionization, referred to as the '$\alpha$-discharge regime.' The long operational lifetime of GDTs in this regime has been validated in our previous high-power limiter and switch designs \cite{Semnani:APL2014, a6semnani2016}.

\subsection{Frequency Tuning}
\begin{figure}[!]
\centering
\includegraphics[width=0.95\linewidth]{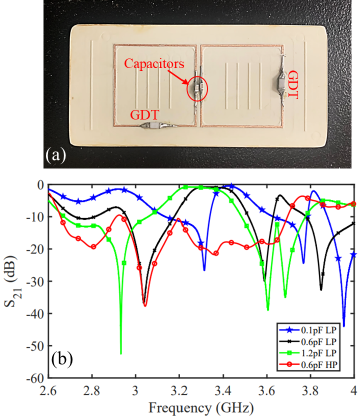}
\caption{(a) The fabricated prototype for frequency-tunable EIT limiter with the tuning capacitors on the vertical arms at the center of the cell. (b) The transmission and reflection responses of the tunable limiter under low (1 W) and high (100 W) power scenarios.}
\label{Fig6: Tun}
\end{figure}

Being frequency tunable to adapt the passband to various applications is an important feature, and this tunability can be implemented in the proposed EIT limiter. SRRs are well-known for their frequency tunability by controlling their gap capacitance. However, the coupling of the resonators makes this tuning scheme challenging in the present scenario. To realize tunable operation for the passband window, high Q capacitors, ranging from 0.1 to 1.2 pF, are introduced on the coupled edges of the resonators, as shown in Fig. \ref{Fig6: Tun}(a). These spots are optimal as they have a low $|E|$-field, as seen in Fig.~\ref{Fig2: S22}(a), which keeps the capacitors safe even under high input powers.

The simulated and measured transmission results for low-power conditions are presented in Fig. \ref{Fig6: Tun}(b). With the capacitance range mentioned above, the passband center frequency can be tuned from 3.1 to 3.5 GHz, with an overall band of 400 MHz. In simulations, the insertion loss of 0.15 to 0.25 dB is observed, while the loss of 0.43 to 0.76 dB is measured. The experimental results show a slight deviation from the simulations mainly due to the requirement for high-precision placement of the capacitive elements. Under a high input power of 50 dBm (100 W), the stopband with an isolation of nearly 20 dB is achieved in the case of 0.6 pF. For the sake of brevity, only o.6 pF results are presented. However, good isolation was achieved for the entire tuning range under high-power conditions.

\subsection{Limiting Threshold Power Tuning}
Various systems require different thresholds for microwave exposure. To ensure the versatility of protection devices, it is essential to adjust the power level at which the limiter device should be turned on and switched to protection mode. With the current topology, a DC biasing scheme is implemented to achieve this power threshold tuning feature. By applying an external DC voltage lower than the breakdown voltage of the GDTs, the microwave breakdown can happen at lower powers because of the pre-ionization phenomenon, which generates more seed electrons and thus enables the microwave breakdown threshold tuning.

\begin{figure}[!]
\centering
\includegraphics[width=0.95\linewidth]{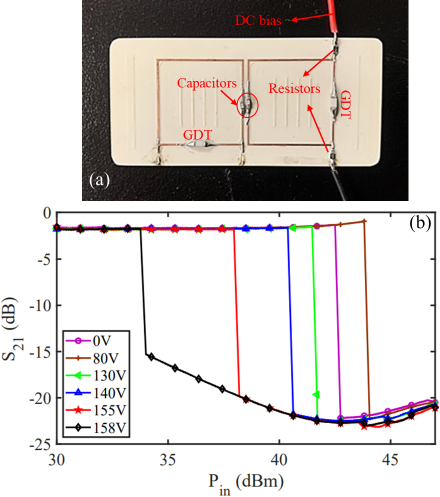}
\caption{(a) The fabricated EIT-based plasma limiter device with the DC threshold tuning scheme. (b) The measured transmission response is in an input power sweep mode at a frequency of 3.3 GHz, utilizing various DC voltages to control the microwave breakdown threshold point.}
\label{Fig7: Thre}
\end{figure}

To experimentally prove this concept, a DC bias is applied to the Littelfuse SE140 GDT in the frequency-tunable limiter device with 0.6 pF capacitors, as seen in Fig. \ref{Fig7: Thre}(a). With the 140 V DC breakdown voltage for this GDT, the threshold tuning DC voltage should be maintained below 140 V to avoid DC breakdown, which is necessary to keep the limiter's low insertion loss in the OFF mode and prevent the lifetime issues of DC discharges \cite{Semnani:freq2013,Semnani:APL2014}. In this configuration, the tuning capacitors in the SRR vertical arms also function as DC block ones, avoiding DC leakage between the two terminals. Additional DC block capacitors would be required at both ends of the GDT to implement the DC bias scheme on the static limiter. To ensure proper DC-RF isolation, DC bias lines are connected through 62 k$\Omega$ resistors. The introduced resistors will incur an additional voltage drop, increasing the threshold DC requirement to 160 V. Moreover, the additional resistors will affect the resonator quality factor, resulting in a slightly higher loss.

The measured transmission coefficients against the input power at 3.3 GHz for the threshold power tuning are presented in Fig. \ref{Fig7: Thre}(b). As the DC voltage increases from 0 to 158 V, the microwave breakdown power is reduced from 43 dBm (20 W) to 34 dBm (2.5 W), showing 9 dB of threshold power tuning. The limiter isolation is proportional to the input power owing to the nonlinear dependence of the electron density and plasma sustaining power and is always more than 15 dB. The higher OFF-mode insertion loss in the power-tunable limiter is due to the employed biasing lines, which had to be taken out of the waveguide structure, resulting in a gap, and also due to the additional resistor and capacitor elements for DC coupling and RF isolation.

The breakdown power threshold initially increases with the applied DC bias up to about 80 V and starts decreasing as the DC bias voltage is further increased. This effect can be explained by the increase in the drift and diffusion of electrons with the application of an external DC voltage, initially causing higher losses of electrons. As the applied voltage further increases, the strong DC field enhances the electron-impact ionization and eventually overcomes the losses through the drift and diffusion of electrons. As a result, the pronounced ionization will require less microwave power for breakdown, lowering the threshold.

\subsection{Wideband EIT Limiters}
\begin{figure}[!]
\centering
\includegraphics[width=0.95\linewidth]{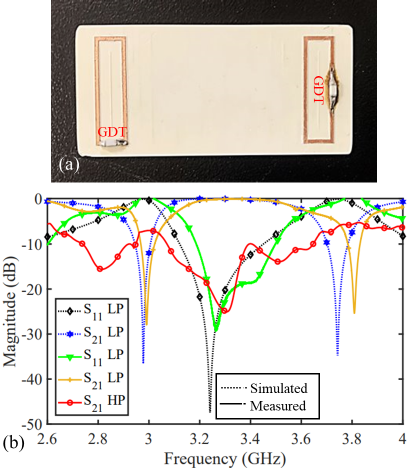}
\caption{The fabricated prototype for a wideband EIT-based limiter, with \textit{l} = 72.14 mm, \textit{w} = 34.04 mm, $w_1$ = 1 mm, \textit{g} = 1 mm, $l_1$ = 22 mm, and $l_2$ = 22 mm. (b) Comparison between simulated and measured transmission and reflection coefficients of the wideband limiter under low power (LP) of 0 dBm (1 W) and high power of 50 dBm (100 W).}
\label{Fig8: wide}
\end{figure}

Operating bandwidth is an important parameter in any RF system, and it can be widely controlled in the proposed EIT-based topology. As the two resonators are moved away from each other, the inter-resonator coupling is decreased, and the EIT bandwidth increases. Fig.\ref{Fig8: wide} shows a sample wideband design with the dimensions provided in the caption. In the experiment, a 15\% passband of 3.11 to 3.63 GHz with a very low insertion loss of 0.08 is achieved, which is in excellent agreement with the simulation results. At high input power of 50 dBm (100 W), an isolation of more than 13 dB is observed throughout the entire passband. There is typically a trade-off between bandwidth and isolation. To achieve both, cascaded limiters, as presented in \cite{vander2020plasma}, are typically used. The operating bandwidth of this limiter can even be tuned electronically by implementing a varactor-tunable inter-resonator coupling structure, which is beyond the scope of this work.

\subsection{Group Delay}
\begin{figure}[!]
\centering
\includegraphics[width=0.95\linewidth]{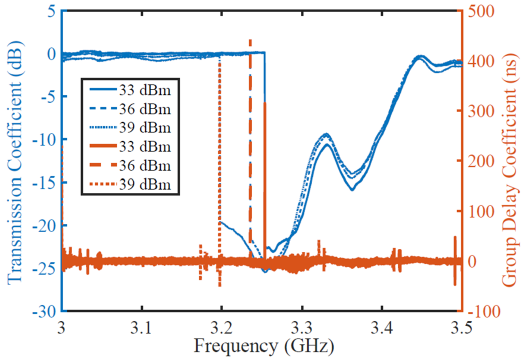}
\caption{Measured group delay along with transmission response normalized to the device OFF-mode performance for the 1.2-pF tunable limiter at various input power levels. A sharp delay is observed at the ignition, while a higher power can ignite the plasma at a lower frequency, i.e., further from the resonant frequency.}
\label{Fig9: gd}
\end{figure}

The transient plasma formation time for the tunable limiter device with 1.2 pF capacitors is evaluated at various input power levels using the group delay characteristics, as shown in Fig. \ref{Fig9: gd}. The setup is first calibrated with the device inserted into the waveguide under low power, i.e., plasma OFF mode, for frequency from 3 to 3.5 GHz. This calibration compensates for the frequency response of the device over the whole band. High power signals are provided to evaluate the ON-mode performance, and group delay is measured along with the transmission coefficient. The results indicate that gas breakdown shifts to lower frequencies with increasing input power, demonstrating easier ignition towards the higher frequencies due to fast collisions. Additionally, the time required for the device to reach a steady state after plasma ignition varies with the input power, ranging from 300 to 400 ns. This is consistent with our response time measurements at a single frequency. 

\subsection{Response Time}
To evaluate the pass-to-isolation switching speed, the time-domain response of the proposed limiter is investigated at 3.2 GHz for both low and high-power scenarios using a high-frequency oscilloscope. The employed pulsed microwave waveform has a pulse width of 4 ms with a duty cycle of 33$\%$. Since this response time depends on the microwave power, this measurement is taken at three different power levels of 20, 40, and 100 W. The long employed pulse width provides enough time for plasma to turn on and off in one cycle so that the oscilloscope can capture the times. Additionally, the long pulse width helps avoid any unnecessary distortions of the waveform due to the higher-order harmonics, keeping in view the cutoff frequency of the waveguide. 

\begin{figure}[!]
\centering
\includegraphics[width=0.98\linewidth]{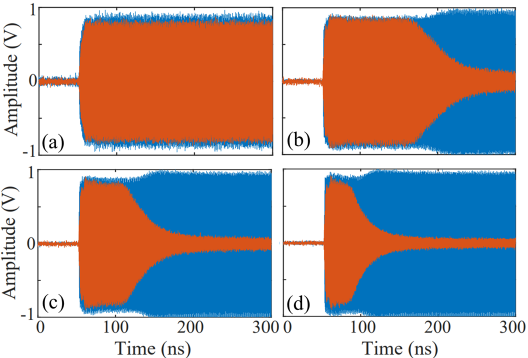}
\caption{The normalized time-domain input (blue) and output (orange) waveforms with a duty cycle of 33\%  and pulse width of 4 ms at (a) low input power of 0 dBm (1 W) with plasma off, and plasma on at (b) 43 dBm (20 W), (c) 46 dBm (40 W), and (d) 50 dBm (100 W).}
\label{Fig10: ss}
\end{figure}

The power-induced protection of the proposed plasma-integrated EIT limiter is observed in Fig.~\ref{Fig10: ss}, indicating a response time in the range of 10s of nanoseconds. As input power increases, plasma quickly reaches a quasi-steady state, lowering the switching speed of the limiter from a passband to protection mode. Specifically, at 100 W of input power, the switching speed is around 60 ns. Notably, the evaluated response time includes a transition period with partially limiting operation. The realized fast switching competes well with semiconductor-based devices but with an additional advantage of much higher power handling and bearing harsh environments.

\begin{table*}[!]
\centering
                \caption{Comparison with the State-of-the-Art  Waveguide Limiter devices}
           \small
           %\vspace{-10pt}
            \renewcommand{\arraystretch}{1.25}
            \begin{center}
%               \begin{tabular}{|p{1.3cm}| p{1.75 cm}| p{1.6cm} | p{1.5cm} | p{1.5cm}  | p{1.5cm} | p{1.25cm}  | p{1.75cm}| p{1.75cm}|}
                \begin{tabular}{c c c c c c c c c c}            \hline        \hline
                   % \vspace{15pt}
& Insertion  & Isolation & Design & No. of  & Tunability  & Response  & Topology & -3dB Band  \\
& loss (dB)  &  (dB) & Complexity & Switches/unit  & Frequency/threshold &  time (ns)& & GHz   \\ \hline 
                                  
                  \cite{a9huang2022}  & $<$ 1 & $>$ 13.5 & 2 metal layers & 3 pin diodes   & No  &  -  & non-resonant & 1.8-2.8\\
 
                   \cite{a10zhang2021} & $<$ 1 & $>$ 14 & 2 metal layers  & 4 pin diodes  & No &  -& HIS$^1$ & 10.5-12\\

                  \cite{a23yang2022} & - & $\approx$10 & 2 metal layers  & 1 pin diode  & No &  -& HIS & 3.4$^2$\\

                  \cite{a72022self} & $<$ 1.3 & $>$ 9 & 3 metal layers  & 4 pin diodes  & No &  -& Non-resonant & 8.5-12.5\\

                  \cite{a8yang2013} & - & $>$ 18 & 1 metal layers  & 2 pin diodes  & No & 250 & Non-resonant & 0.5-1.5\\

                  \cite{a12zhou2021} & $<$ 1  & $>$ 24 & 2 metal layers  & 4 pin diodes  & No & - & double resonance & 2.8\\

                  \cite{a12zhou2021} & $<$ 2.5  & 18  & 1 metal layers  & 4 pin diodes  & No & - & resonant & 2.7\\

                  Static & $<$ 0.26  &  $>$ 26  & 1 metal layers  & 1 GDT$^3$  & yes & $\approx$ 60 & EIT & 2.86-3.08\\
                  Limiter &   &    &   &   &  &  &  & \\

                  Wideband& $<$ 0.08  & $>$ 13  & 1 metal layers  & 1 GDT  & Yes & $\approx$ 60 & EIT & 3.11-3.63\\
                  Limiter &  &   &  &   &  &  &  & \\

                         \hline
                    \hline
                \end{tabular}
                \label{tab1:comp}
            \end{center}

            \footnotesize{$^1$High impedance surfaces. $^2$Single Frequency indicating narrowband. $^3$Practically 0.5 GDT switch as each SSR is equivalent to a unit and only 1 GDT works as a switch per two SSRs; 2nd one is introduced for symmetry purposes.}
            
        \end{table*}

\section{Conclusion}
A novel high-power microwave protection scheme has been developed, employing coupled resonators to induce an electromagnetically induced transparency (EIT) effect, creating a passband within a broad stopband. Upon reaching a specific incident power threshold, a plasma cell integrated into one of the resonators is ignited, disrupting the resonator symmetry and negating the EIT condition, which results in the formation of a stopband to achieve protection. This concept was validated through the implementation of a static limiter operating at 3 GHz, exhibiting a 3-dB passband window of 7.4\%, a wideband limiter with a 3-dB passband of 15\%, and a frequency-tunable limiter with 400 MHz of frequency tuning capability. Additionally, the system demonstrated nearly 10 dB of power threshold tuning through an external DC biasing mechanism. The response time of the limiter was measured to be approximately 60 ns at an input power of 100 W, with a group delay ranging from 15 to 20 ns, contingent on the input power level. In conclusion, the proposed solution leverages the physics-based EIT phenomenon for the first time to demonstrate a front-end microwave limiter, surpassing existing devices in several key aspects.

\bibliographystyle{IEEEtran}

\bibliography{ref}

%\newpage

%If you have an EPS/PDF photo (graphicx package needed), extra braces are needed around the contents of the optional argument to biography to prevent  the LaTeX parser from getting confused when it sees the complicated  $\backslash${\tt{includegraphics}} command within an optional argument. (You can create  your own custom macro containing the $\backslash${\tt{includegraphics}} command to make things  simpler here.)
 

%\bf{If you include a photo:}\vspace{-33pt}
%\begin{IEEEbiography}[{\includegraphics[width=1in,height=1.25in,clip,keepaspectratio]{fig1}}]{Michael Shell}
%Use $\backslash${\tt{begin\{IEEEbiography\}}} and then for the 1st argument use $\backslash${\tt{includegraphics}} to declare and link the author photo.
%Use the author name as the 3rd argument followed by the biography text.
%\end{IEEEbiography}


%bf{If you will not include a photo:}\vspace{-33pt}

%\vfill

\end{document}